\documentclass[11pt]{article}
\usepackage[utf8]{inputenc}
\pdfoutput=1
\usepackage{amssymb,amsmath,mathrsfs,enumerate}
\usepackage{graphicx,rotate,multicol}
\usepackage{float}
\usepackage{tocloft}
\usepackage{subfig}
\usepackage[margin=10pt,labelfont=bf]{caption}
\usepackage{cite}
\usepackage{soul}
\usepackage[colorlinks=true,
            linkcolor=blue,
            urlcolor=blue,
            citecolor=teal]{hyperref}
\usepackage{multirow}
 \usepackage{placeins}

\makeatletter
\let\Hy@linktoc\Hy@linktoc@page
\makeatother

\usepackage{color}
\definecolor{ourcolor}{rgb}{0.7, 0.25, 0.05}
\usepackage{tikz,braket}
\long\def\rpl#1!!#2!!{\textcolor{red}{#1} \textcolor{blue}{#2}}

\def \order(#1){{\mathcal O} \left(#1 \right)}

\evensidemargin=\oddsidemargin
\allowdisplaybreaks

\definecolor{lime}{HTML}{A6CE39}
\DeclareRobustCommand{\orcidicon}{\hspace{-1mm}
	\begin{tikzpicture}
	\draw[lime, fill=lime] (0,0) 
	circle [radius=0.16] 
	node[white] {{\fontfamily{qag}\selectfont \tiny \,ID}};
	\draw[white, fill=white] (-0.0525,0.095) 
	circle [radius=0.007];
	\end{tikzpicture}
	\hspace{-3mm}
}
\foreach \x in {A, ..., Z}{\expandafter\xdef\csname orcid\x\endcsname{\noexpand\href{https://orcid.org/\csname orcidauthor\x\endcsname}
			{\noexpand\orcidicon}}
}

\begin{document}

\title{\color{black}{\bf Cosmic-ray boosted dark matter in Xe-based direct detection experiments}}

\author {\bf Tarak Nath Maity\orcidA{},$^{a}$\footnote{tarak.maity.physics@gmail.com} 
\hspace{4pt}  Ranjan Laha\orcidB{}$^{a}$\footnote{ranjanlaha@iisc.ac.in}
\\[10pt]
\small\em $^a$Centre for High Energy Physics, Indian Institute of Science, C.\,V.\,Raman Avenue, Bengaluru 560012, India
}
\date{}

\maketitle

\begin{abstract}
LUX-ZEPLIN (LZ) collaboration has achieved the strongest constraint on weak-scale dark matter (DM)-nucleon spin-independent (SI) scattering cross section in a large region of parameter space. In this paper, we take a complementary approach and study the prospect of detecting cosmic-ray boosted sub-GeV DM in LZ. In the absence of a signal for DM, we improve upon the previous constraints by a factor of $\sim 2$ using the LZ result for some regions of the parameter space. We also show that upcoming XENONnT and future Darwin experiments will be sensitive to cross sections smaller by factors of $\sim 3$ and $\sim 10$ compared to the current LZ limit, respectively.
\end{abstract}

\section{Introduction}
\label{sec:intro}
A multitude of cosmological and astrophysical observations indicate that the biggest slice ($\sim 85\%$) of the matter density of the Universe is made up of DM\,\cite{Green:2021jrr, Planck:2018vyg, Lisanti:2016jxe}. While the presence of DM is revealed through gravitational observations, its true nature is yet to be known. Typically it is assumed that DM might be a particle in nature, and depending on the nature of the particle, the allowed DM mass range varies. Additionally, it is phenomenologically interesting to have an interaction between different  Standard Model (SM) states and DM. This relation is also common in a myriad of well-motivated particle physics models \cite{Arcadi:2017kky, Bauer:2017qwy, Adams:2022pbo, 10.21468/SciPostPhysLectNotes.43}. Many ongoing and upcoming searches are specifically looking for this connection \cite{Buchmueller:2017qhf, Krnjaic:2022ozp, Batell:2022dpx, PerezdelosHeros:2020qyt, Slatyer:2021qgc, Cooley:2021rws, Kahn:2021ttr}.

Direct detection (DD) experiments look for the recoil of SM states through its scattering with ambient DM particles, and is mostly relevant for weakly interacting massive particles (WIMPs) searches. One type of DD experiments hunt for recoil of the target nucleus, kept in underground laboratories\,\cite{SuperCDMS:2015eex, SuperCDMS:2018mne, PICO:2017tgi, DarkSide:2022dhx, DarkSide-50:2022qzh, LUX:2018akb, XENON:2018voc, PandaX-4T:2021bab}. Among various target materials, xenon stands out to be quite beneficial due to its properties like shelf shielding, higher mass number, inert chemical nature, and others. Interestingly, Xe target experiments continue to set the leading limits in large regions of DM parameter space \cite{LUX:2018akb, XENON:2018voc, PandaX-4T:2021bab}. The main target material in a two-phase time projection chamber (TPC) is liquid xenon (LXe). The possible DM interaction with LXe is detected through light yields (S1) and charge yields (S2). Combining the S1 and S2 signal topologies, it is possible to reconstruct the event's three-dimensional position and efficiently discriminates between nuclear recoil (NR) and electron recoil (ER) signatures, etc. The NR backgrounds arise from neutrons, neutrino-nucleus interactions, etc., whereas the ER background arise and from $\beta$-decays, $\gamma$ produced by radioactivity, neutrino-electron interactions, etc. Experiments like Xenon, LZ, and PandaX are exploring possible DM events in the presence of these backgrounds.

Recently, LZ collaboration has published its first result \cite{LUX-ZEPLIN:2022qhg}. The experiment is situated at $4850$ ft underground in the Davis Cavern at the Sanford Underground Research Facility (SURF) in Lead, South Dakota, USA. The total mass of LXe is $10$ t, out of which only the inner fiducial $5.5$ t is used for DM searches to reduce the backgrounds. With $60$ live days of data, LZ has reached the current strongest constraint $6 \times 10^{-48} {\rm cm^2}$ at DM mass $30$ GeV. Compared to the previous strongest bound, this is $6.7$ and  $1.7$ times better at DM mass $\sim 30$ GeV and $\sim 1000$ GeV, respectively.

While the LZ  result focuses on the searches for WIMP-like DM, in this paper, we take a complementary approach to investigate scenarios of sub-GeV DM interacting with nucleons via spin-independent (SI) interactions. Non-relativistic sub-GeV DM, typically moving with velocity $\sim 10^{-3}$, will not be able to impart enough energy to produce an observable NR in the LZ experiment. However, an energetic sub-GeV DM may produce sufficiently large NR. One of the simplest ways to produce such boosted DM is to consider the interaction between high-energy cosmic-rays (CRs) and DM, known as  CR boosted DM (CRDM), proposed for the first time in Ref.\,\cite{Bringmann:2018cvk} for nuclear scattering and Ref.\,\cite{Ema:2018bih} for electron scattering. Further this technique has received considerable attention\,\cite{Kouvaris:2015nsa, An:2017ojc, Yin:2018yjn, Alvey:2019zaa, Cappiello:2019qsw, Dent:2019krz, Krnjaic:2019dzc, Bondarenko:2019vrb, Wang:2019jtk, Guo:2020drq, Ge:2020yuf, Jho:2020sku, Jaeckel:2020oet, Cho:2020mnc, Guo:2020oum, Dent:2020syp, Ema:2020ulo, Dent:2020qev, PROSPECT:2021awi, Flambaum:2020xxo, Jho:2021rmn, Emken:2021lgc, Das:2021lcr, Bell:2021xff, Chao:2021orr, An:2021qdl, Bramante:2021dyx, Ghosh:2021vkt, Feng:2021hyz, Wang:2021nbf, Xia:2021vbz, Wang:2021jic, CDEX:2022fig, Granelli:2022ysi, Xia:2022tid, Bardhan:2022ywd, Cline:2022qld, Ferrer:2022kei, Alvey:2022pad, Li:2022dqa, Kolesova:2022kvq}. These boosted DM particles reach the underground detector with much higher energy which helps to overcome the energy threshold although with much lower flux. Even with this lowered flux, it is possible to probe new regions of DM-nucleon scattering cross-section, since the bounds for sub-GeV DM using other techniques are weak. The paradigm of CRDM premises only on the assumption of DM-nuclear interactions, which is also true for many DD experiments. A large class of particle physics models predicts such interaction for sub-GeV DM\,\cite{Boehm:2003hm, Pospelov:2007mp, Izaguirre:2015yja, Knapen:2017xzo, Boehm:2020wbt, Elor:2021swj}.

Knowledge of the CR spectrum is an important ingredient in computing CRDM flux. The direct CR flux measurements (PAMELA\,\cite{PAMELA:2011mvy}, AMS-02\,\cite{AMS:2015tnn, AMS:2015azc}, CREAM-I\,\cite{Yoon:2011aa}, etc.) are done with balloons and satellite detectors near the top or outside the atmosphere. This has been used as input CR flux in Ref.\,\cite{Bringmann:2018cvk}. However above $~100$ TeV CR fluxes are small hence direct measurements are not a feasible choice. In this case, CR is measured indirectly through the air shower induced by it. We utilize the parametric fit of CR flux measurement (obtained by combining direct and indirect CR flux measurements) given in Ref.\,\cite{Gaisser:2013bla} as the input CR flux. Then we explore the signature of the CR-induced DM in the LZ experiment. We find a factor $\sim 2$ improvement compared to previous limit of XENON1T near DM mass $\sim 1$ MeV. We also present the projections of the upcoming XENONnT, LZ, and Darwin in probing the DM-nucleon cross-section for sub-GeV DM. We find that there can be a factor $\sim 10$ improvement for Darwin compared to the current LZ limits.

The paper is organized as follows. In Sec.\,\ref{sec:CRDM}, we briefly sketch the CRDM framework. In Sec.\,\ref{sec:CRDMDD}, we present limits from LZ and future xenon-based experiments. We conclude in Sec.\,\ref{sec:con}.
%
%
%
%
\section{Overview of CRDM}
\label{sec:CRDM}
Let us consider a DM particle ($\chi$) of mass $m_\chi$, scattering with a CR particle of mass $m_i$. After scattering, the CR induced DM flux is\,\cite{Bringmann:2018cvk}
\begin{equation}
\frac{d\phi_\chi}{dT_\chi} = D_{\rm eff} \frac{\rho_\chi^{\rm local}}{m_\chi} \sum_i \sigma_{\chi i} G_i^2(2 m_\chi T_\chi) \int_{T_i^{\rm min}}^{\infty} dT_i \frac{1}{T_{\chi}^{\rm max}(T_i)}  \frac{d\phi_i^{\rm CR}}{dT_i},
\label{eq:DMFlux}
\end{equation}
where $T_{\chi}$ and $T_i$ are the DM and CR kinetic energies respectively. The effective distance, $D_{\rm eff}$, depends on the distance to which DM flux is integrated. The local DM density is denoted by $\rho_\chi^{\rm local}$, fixed to $0.3\,{\rm GeV/cm}^3$. In the sum, we have included the contributions of p, He, C, O, and Fe. DM-nucleus scattering cross section is represented by $\sigma_{\chi i}$ and $G_i^2(2 m_\chi T_\chi)$ is the nuclear form factor. The differential CR flux is represented by $d\phi_i^{\rm CR}/dT_i$. The minimum CR energy required to produce a DM of kinetic energy $T_{\chi}$ is
\begin{equation}
\label{eq:Timin}
T_i^{\rm min} = \left( \frac{T_\chi}{2}-m_i \right)\left(1 \pm \sqrt{1+\frac{2 T_\chi}{m_\chi} \frac{(m_i+m_\chi)^2}{(2 m_i-T_\chi)^2}} \right),
\end{equation}
with $+$ and $-$ sign applicable to $T_\chi > 2 m_i$ and $T_\chi < 2 m_i$ respectively. The maximum kinetic energy transferred to DM by the CR DM collision is given by
\begin{equation}
\label{eq:Tchimax}
T_\chi^{\rm max} = \frac{2 m_\chi\left(T_i^2+2 m_i T_i\right)}{ 2 m_\chi T_i+(m_i+m_\chi)^2}
\end{equation}
The effective distance is
\begin{equation}
\label{eq:deff}
D_{\rm eff} = \frac{1}{\rho_\chi^{\rm local}} \int \frac{d \Omega}{4 \pi} \int_{\rm los} \rho_{\chi} d\ell,
\end{equation}
where $\rho_{\chi}$ is the Milky-Way (MW) DM density profile under consideration.  The angular region of the integration is represented by $d\Omega$. For traditional Navarro-Frenk-White DM density profile \cite{Navarro:1996gj} (with parameters taken from Ref.\,\cite{Maity:2021umk}), the effective distance (given in Eq.\,\eqref{eq:deff}) turns out to be $\sim 1$ kpc and $\sim 10$ kpc for integrating up to $1$ kpc and $10$ kpc around the Sun respectively\,\cite{Bringmann:2018cvk}. In our numerical calculation we fix $D_{\rm eff}$ to $10$ kpc.  We have used the Global fit model presented in Ref.\,\cite{Gaisser:2013bla} inspired by various CR measurement. This CR spectra retain the various spectral features which arise due to the contribution from different possible CR sources. We do not consider spatial-dependent CR flux; including them, will strengthen our limits and sensitivities by factor of $\sim 3$\,\cite{Xia:2021vbz}, hence our results are conservative.

The CRDM particles will also interact with the nuclei while traversing from the top of the atmosphere to the underground detector. CRDM will lose its energy for a reasonably large cross-section due to scattering with these nuclei. The subsequent attenuation of CRDM flux has been studied in Refs.\,\cite{Bringmann:2018cvk, Xia:2020apm, Xia:2021vbz}.\footnote{See Refs.\,\cite{Davis:2017noy, Kavanagh:2017cru, Hooper:2018bfw, Emken:2018run, Emken:2019tni} for attenuation of non-boosted DM with large DM-SM scattering cross sections.} We calculate the DM kinetic energy at a depth $z$, $T_{\chi}^z$, utilizing\,\cite{Alvey:2022pad} 
\begin{equation}
\label{eq:att}
\frac{dT_{\chi}^z}{d z}= - \sum_j \frac{\rho}{m_j} \int_0^{\omega_{\chi}^{\rm max}} d \omega_{\chi} \frac{d \sigma_{\chi j}}{d \omega_{\chi}} \omega_{\chi},  
\end{equation}
where $\rho$ is the average mass density of the medium, $m_j$ is the mass of the target nucleus and $\omega_{\chi}$ is the energy loss of DM particles due to collision. For elastic scattering, the maximum energy loss, $\omega_{\chi}^{\rm max}$, can be read off from Eq. \eqref{eq:Tchimax}. The average mass density of the Earth's crust is assumed to be $2.7 \, {\rm gm/cm}^3$. The corresponding weightage for each nuclei is adapted from \cite{Bringmann:2022vra}. Following DarkSUSY\,\cite{Alvey:2022pad, Bringmann:2022vra},  we have also included inelastic DM-nucleus scattering by doing a correspondence with neutrino-nucleus scattering. The neutrino-nucleus  inelastic scattering (i.e., quasi elastic, deep inelastic scattering, $\Delta$ and other hadronic resonances) cross section is calculated using \textsc{GiBUU}\,\cite{Buss:2011mx,gibuu}. The effect of form factor in the differential cross section, has also been included. The form factor for different element are obtained from \cite{DeVries:1987atn}. We have not included DM-nucleus scattering in the atmosphere, owing to the low atmospheric density ($\lesssim 1.2 \times 10^{-3} \, {\rm gm/cm}^3$)\,\cite{Emken:2018run}, as it will have a negligible impact, for the considered cross section.

The CRDM particles reach the underground detector and collide with the nucleus of the target material to produce an observable recoil. The differential recoil rate per unit target nucleus is\,\cite{Bringmann:2018cvk}
\begin{equation}
\label{eq:dRateGen}
\frac{d R}{dE_{\rm N}} = \sigma_{\chi N} \, G_N^2(2 m_N E_{\rm N}) \int_{T_{\chi}(T_{\chi}^{z,{\rm min}})}^{\infty} \frac{dT_{\chi}}{E_N^{\rm max}(T_{\chi}(z))} \frac{d\phi_\chi}{dT_\chi},
\end{equation}
where $E_N$ is the nuclear recoil energy. The maximum nuclear recoil ($E_N^{\rm max}$) and the required minimum DM kinetic energy at detector location ($T_{\chi}^{z,{\rm min}}$) can be obtained by treating nucleus and DM as target and incident particle, respectively, in Eqs.\,\eqref{eq:Timin} and \eqref{eq:Tchimax}. In Eq.\,\eqref{eq:dRateGen}, the lower limit of the integration, $T_{\chi}(T_{\chi}^{z,{\rm min}})$ which is the required DM kinetic energy at the surface of the earth so that DM partcles reaches the underground detector with $T_{\chi}^{z,{\rm min}}$, can be obtained utilizing the solution of Eq.\,\eqref{eq:att}. We concentrate on isotropic, elastic, spin-independent (SI) DM-nuclear scattering throughout the paper. In this case, the nuclear level cross-section is related to the nucleon level through
\begin{equation}
\label{eq:Nucleartonucleon}
\sigma_{\chi N} = \sigma_{\chi}^{\rm SI} A^2 \left(\frac{m_N(m_{\chi}+m_p)}{m_p(m_{\chi} +m_N)}\right)^2.
\end{equation}
Here $A,~m_N,$ and $m_p$ are the mass number, nuclear mass, and proton mass respectively. The SI DM-nucleon cross section is denoted by $\sigma_{\chi}^{\rm SI}$.

\section{CRDM at LZ and future xenon based experiments}
\label{sec:CRDMDD}
The search for CRDM in liquid xenon detectors have been previously studied in the literature\,\cite{Bringmann:2018cvk,Xia:2021vbz,PandaX-II:2021kai,Alvey:2022pad}. This section focuses on probing CRDM using the latest LZ results. We also present the projected sensitivities of XENONnT, LZ (future data set), and Darwin experiments.  

\subsection{CRDM at LZ}
\label{subsec:CRDMLZ}

\begin{figure}[t]
	\begin{center}
		\includegraphics[scale=0.28]{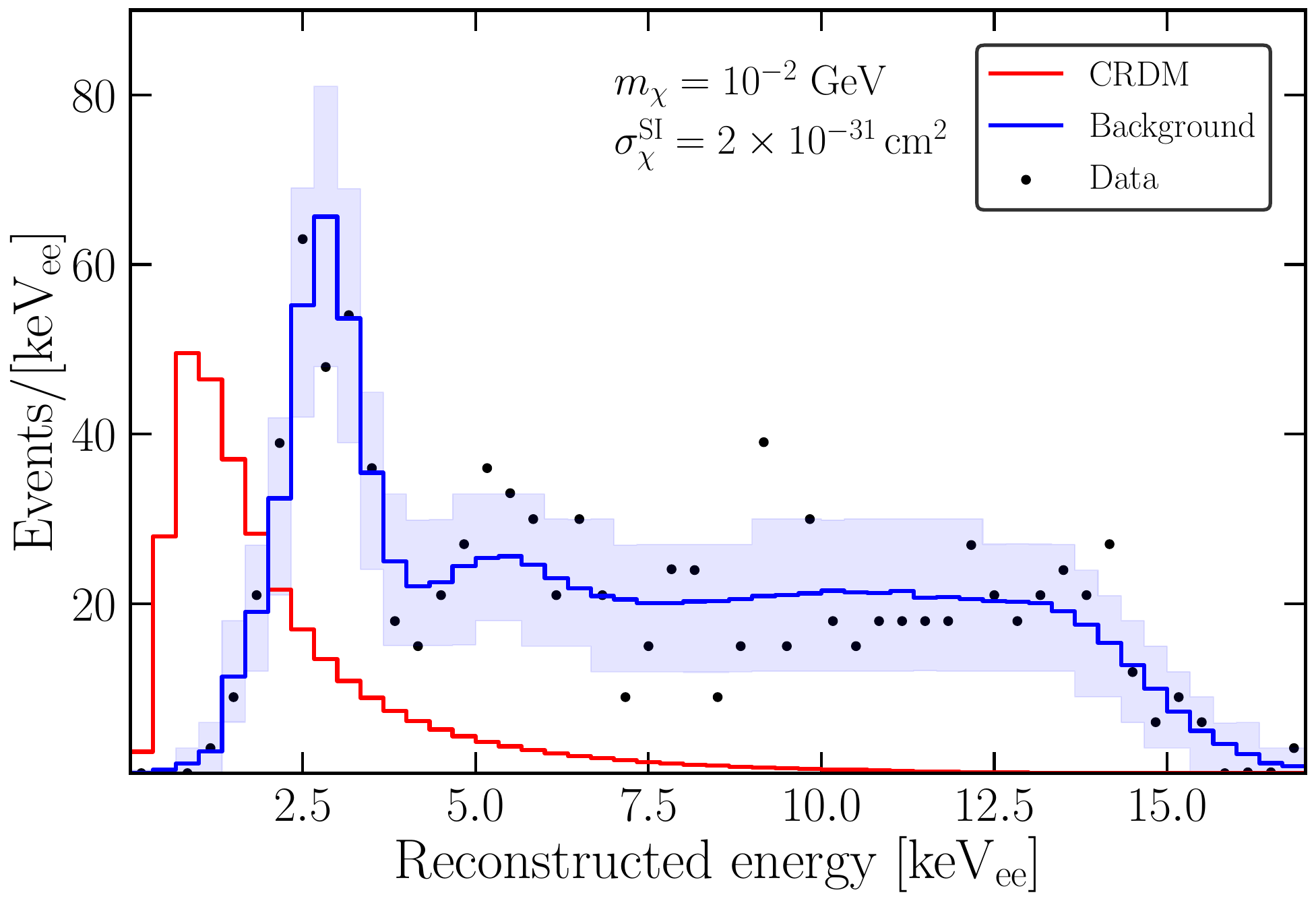}
		\caption{Differential event rate against the electron equivalent reconstructed energy with $5.5$\,t Xe and 60 days exposure. The solid blue line (light blue shaded band) shows the central value of the background (the systematic and statistical uncertainties) as reported in Ref.\,\cite{LUX-ZEPLIN:2022qhg}. The black points represent the data. The solid red line represents expected signal events (including signal efficiency) from CRDM assuming the DM parameters as mentioned in the figure.   }
		\label{fig:dRdEeeLZ}
	\end{center}
\end{figure}

Here we explore prospect of detecting CRDM in light of the recent LZ result.\footnote{See Refs.\,\cite{Barman:2022jdg, AtzoriCorona:2022jeb, A:2022acy} for other beyond SM searches using LZ data.} In absence of any excess events over the background within the observed energy region we can constrain SI and velocity independent DM-nucleon scattering for CRDM. In the higher recoil energy region, where LZ has not presented their experimental data, the CRDM flux would be further suppressed (for the DM-nuclear interaction that we have assumed in this work), hence there is a lesser possibility of observing the signal. The LZ collaboration has presented the differential recoil rate with respect to electron recoil equivalent energy. The nuclear recoil energy is related to equivalent electron recoil energy by\,\cite{Essig:2018tss}
%
\begin{equation}
\label{eq:quenching}
E_{\rm ee} = Y(E_N) E_N,
\end{equation}
where $Y(E_N)$ is the quenching factor. We use the theoretical model of Lindhard to estimate the quenching factor\,\cite{lindhard1963integral}
\begin{equation}
\label{eq:QF}
Y(E_N)= \frac{k \, (3 \, \epsilon^{0.15} + 0.7 \epsilon^{0.6} +\epsilon)}{1+ k \, (3 \, \epsilon^{0.15} + 0.7 \epsilon^{0.6} +\epsilon)}\,.
\end{equation}
Here $\epsilon=11.5 \, Z^{-7/3} (E_N/{\rm keV})$ and $Z$ is the atomic number. In our calculation, we fix $k$ to 0.145, nearly reproducing LZ electron recoil energy given a nuclear recoil energy (e.g., with $E_N=15\,$keV the difference is approximately $1\%$). Using Eqs.\,\eqref{eq:dRateGen} and \eqref{eq:quenching}, the differential recoil rate with respect to $E_{\rm ee}$ can be expressed as\,\cite{Essig:2018tss}
\begin{equation}
\label{eq:dRdEee}
\frac{dR}{dE_{\rm ee}} = \frac{d R}{dE_{\rm N}} \frac{1}{Y(E_N)+E_N \frac{dY(E_N)}{dE_N}}\,.
\end{equation}
With Xe as the target material, $5.5$ ton fiducial mass, and $60$ live-days, the corresponding CRDM differential recoil rate for LZ is shown in  Fig.\,\ref{fig:dRdEeeLZ} by the solid red line. The CRDM recoil rate is shown for $m_{\chi} = 10^{-2}$ GeV and $\sigma_{\chi}^{\rm SI} = 2 \times 10^{-31}\, {\rm cm}^{2}$. Following Ref.\,\cite{LUX-ZEPLIN:2022qhg}, we include the signal efficiency, which leads to the fall in the event rate in the lowest energy bins.  We have also displayed the expected total background by the solid blue line. We show the associated statistical and systematic uncertainties by the light blue shading. The black points represent the data. Clearly, $m_{\chi} = 10^{-2}$ GeV and $\sigma_{\chi}^{\rm SI} = 2 \times 10^{-31}\, {\rm cm}^{2}$ is ruled out by the recent LZ result. 
\begin{figure}[t]
	\begin{center}
		\includegraphics[scale=0.3]{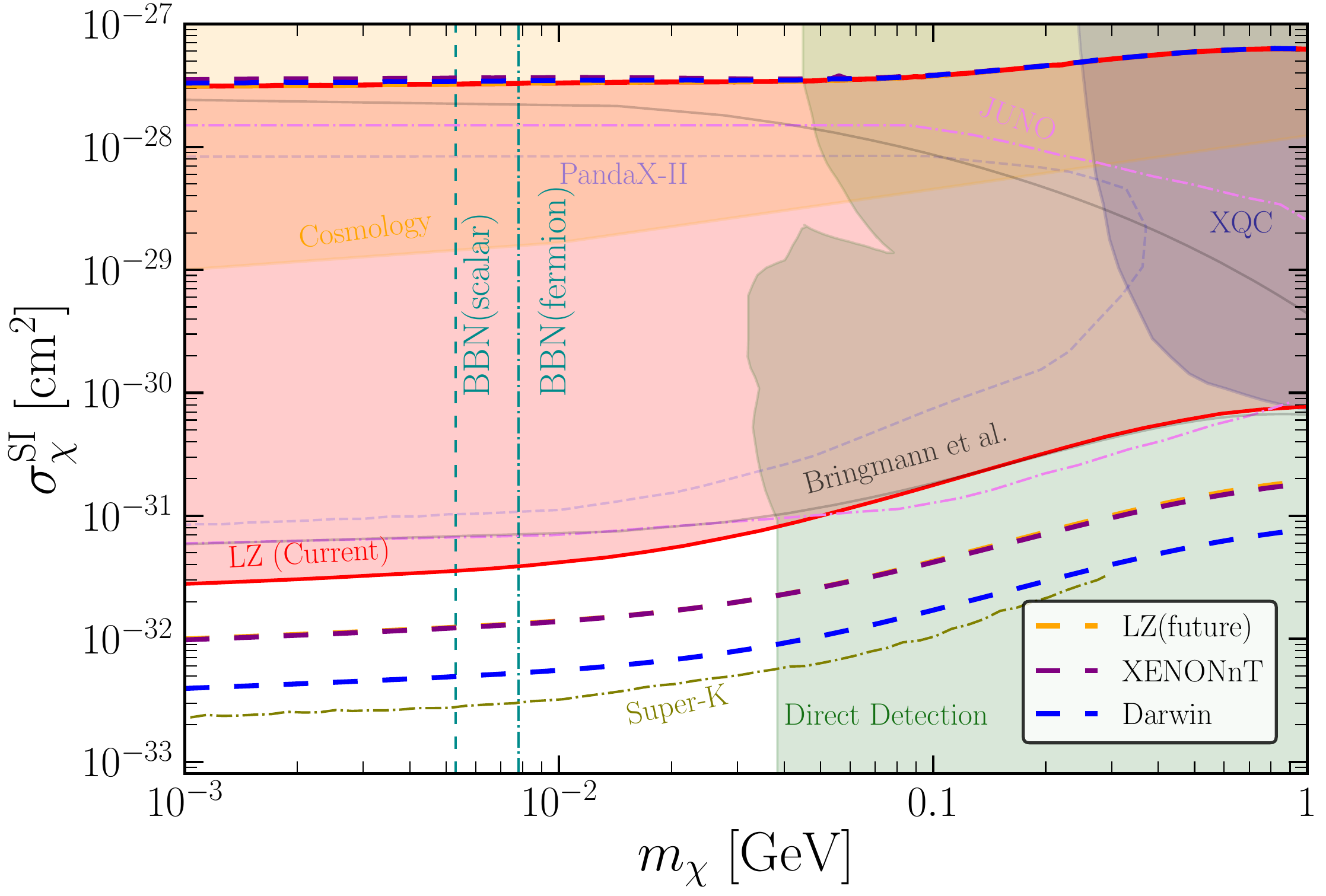}
		\caption{Bounds on SI DM-nucleon scattering cross-section. The light red shaded region displays the current LZ bound on CRDM derived in his work. The constraints reported in Refs.\cite{Bringmann:2018cvk}, and \cite{PandaX-II:2021kai} are denoted by Bringmann et al.\,and PandaX-II and are shown by light  black solid and light blue dashed lines, respectively. Our current LZ limit is complementary to these due to the differences in the input CR fluxes, see the text for details. Using the same cosmic ray flux as in Ref.\,\cite{Bringmann:2018cvk} our calculated limit and sensitivities will further improve by factor $\sim 2.$  We display the recently reported Super-Kamionkande result limit by olive dot-dashed line\,\cite{Super-Kamiokande:2022ncz}.  We also show the future LZ, XENONnT, and Darwin sensitivities on CRDM by the dashed orange, purple, and blue lines, respectively; the future LZ and XENONnT sensitivities overlap.  Other DD, cosmological and XQC limits are also shown by light green, blue, and orange shading, respectively. The future JUNO limit is shown by dot dashed violet line\,\cite{Cappiello:2019qsw}. The BBN limit on complex scalar and Dirac fermion DM are shown by dark cyan dashed and dot-dashed vertical lines. We do not show our constraints for DM masses $\geq 1$ GeV since the parameter space is robustly ruled out by other experiments.}
		\label{fig:limit}
	\end{center}
\end{figure} 

In our statistical analysis of LZ data, we use the $\chi^2$ for Poisson distributed data\cite{Baker:1983tu, ParticleDataGroup:2020ssz}
\begin{equation}
\label{eq:chisquare}
\chi^2 = 2 \sum_{i=1}^{51} N_i^{\rm sig} + B_i -D_i + D_i\, {\rm ln}\left( \frac{D_i}{N_i^{\rm sig} + B_i}\right), 
\end{equation}  
where $N_i^{\rm sig}$, $B_i$, and $D_i$ are the CRDM signal, background events, and data points in the $i^{\rm th}$ bin. In our numerical calculations, we include the central value of the expected background events and $D_i$ reported by LZ. Inclusion of systematic and statistical uncertainties lead to factor $\sim 2$ change in our LZ lower limit. The expected CRDM events can be obtained from Eq.\,\eqref{eq:dRdEee}. The signal events $N_i^{\rm sig}$ is mainly regulated by $m_{\chi}$ and $\sigma_{\chi}^{\rm SI}$, can be evaluated by integrating Eq.\,\eqref{eq:dRdEee} over the corresponding electron recoil energy bin. The $90\%$ confidence level (CL) limit is obtained when $\Delta \chi^2$ ($ = \chi^2 - \chi^2_{\rm min}$) becomes $2.71$. The $\chi^2_{\rm min}$ can be obtained by effectively setting $N_i^{\rm sig}$ to zero in Eq.\,\eqref{eq:chisquare}, since there is no statistically significant excess over the background models.  We show our exclusion bound in Fig.\,\ref{fig:limit}. We vary SI DM-nucleon cross section for a fixed DM mass until the required value of $\Delta \chi^2$ is achieved. We have followed this procedure to estimate the lower bound on the DM-nucleon cross-section. The upper limit represents the cross section for which CRDM would not be able to produce an observable recoil at DD experiments due to the attenuation of DM through its scattering with Earth's nuclei. The difference in shape of our limit, PandaX-II limit and JUNO sensitivity (mentioned later) is mainly due to the treatment of attenuation. The treatment of form factor during attenuation extend our limit to higher DM mass \cite{Alvey:2022pad}. Like PandaX-II limit, our limit also closes but at around DM mass $10\,$GeV. We have omitted showing these regions since they are robustly ruled out by typical DD experiments.

In Fig.\,\ref{fig:limit}, our obtained limit using the recent LZ result is shown by the red contour, and it rules out the light red shaded regions of the parameter space. Our lower bound on SI DM-nucleon cross-section is a factor of $\sim 2$ stronger than XENON1T bound reported in Ref.\,\cite{Bringmann:2018cvk} (denoted as Bringmann et al.\,in Fig.\,\ref{fig:limit} by the light black solid  line) near the DM mass $\sim 1$ MeV.  Unlike our case, Ref.\,\cite{Bringmann:2018cvk} utilized direct CR flux as the input flux. We note that if we consider direct CR flux (like Ref.\,\,\cite{Bringmann:2018cvk}) and analyze the LZ data; there will be a further factor $\sim 2$ improvement \cite{Xia:2020apm} over XENON1T bound presented in Ref.\,\cite{Bringmann:2018cvk}. The olive dot-dashed line displays recent directional search result of CRDM near the Galatic center by Super-Kamiokande (Super-K) \cite{Super-Kamiokande:2022ncz}. The said Super-K result is an order of magnitude stronger than current LZ limit. We also display the other DD\,\cite{SuperCDMS:2013eoh, DAMIC:2016lrs, NEWS-G:2017pxg, SuperCDMS:2018mne, CRESST:2019jnq, CDEX:2019hzn, XENON:2019zpr, EDELWEISS:2019vjv, EDELWEISS:2022ktt, DarkSide:2022dhx, DarkSide-50:2022qzh}, cosmological\,\cite{Gluscevic:2017ywp, Xu:2018efh, Slatyer:2018aqg, Nadler:2019zrb,  Buen-Abad:2021mvc, Rogers:2021byl}, and XQC\,\cite{Mahdawi:2018euy} constraints by the light green, light orange, and light blue shaded regions, respectively.  Following Ref.\,\cite{Krnjaic:2019dzc}, the Big Bang Nucleosythensis (BBN) limits on complex scalar and Dirac fermion DM are displayed by the dashed and dot-dashed dark cyan lines, respectively. Note that for real scalar DM, BBN observations allow the DM mass to be $\geq 1$ MeV\,\cite{Krnjaic:2019dzc}. Near future data of LZ with larger exposure will certainly be able to probe lower cross-sections in sub-GeV DM mass regime. We discuss this issue in the next section.

\subsection{CRDM at future direct detection experiments}
\label{subsec:CRDMProj}
In this section, we discuss the projections of ongoing liquid xenon detectors and future DD experiments in scrutinizing CRDM. We focus on the sensitivity of XENONnT, future data set of LZ, and Darwin experiments. The sensitivities of these experiments are obtained following the prescription of Ref.\,\cite{Bringmann:2018cvk}. The future limit on SI DM-nucleon scattering of CRDM is evaluated using
\begin{equation}
\label{eq:CRDMproj}
\sigma_{\chi}^{\rm SI} = \frac{2 \kappa v_0}{\sqrt{\pi}} \rho_{\chi}^{\rm local} \left( \frac{m_{\chi} + m_N}{m_{\chi} + m_p} \right)^2 \left(\frac{\sigma_{\rm NRDM}^{\rm SI}}{m_{\rm NRDM}}\right) \left(\int_{E_{N_1}}^{E_{N_2}} dE_N \int_{T_{\chi}(T_{\chi}^{z,{\rm min}})}^{\infty} \frac{dT_{\chi}}{E_N^{\rm max}(T_{\chi}(z))} \frac{d\phi_\chi}{dT_\chi}\right)^{-1},
\end{equation}     
%
%
where the numerical factor $\kappa$ is obtained from\,\cite{Bringmann:2018cvk}. The circular velocity of the Sun ($v_0$) is assumed to be $220$ km/s. The projected limit on the ratio of non-relativistic DM (NRDM) nuclear cross section and mass ($\sigma_{\rm NRDM}^{\rm SI}/m_{\rm NRDM}$) is calculated at NRDM mass $\sim 1$\, TeV.  The experiment's recoil energy regime of operation is denote by $E_{N_1}$ and $E_{N_2}$.
\begin{table}[]
\begin{center}
\begin{tabular}{|c|c|c|c|c|}
\hline
Experiment  & \begin{tabular}[c]{@{}c@{}}$E_{N_1}$\\ (keV)\end{tabular} & \begin{tabular}[c]{@{}c@{}}$E_{N_2}$\\ (keV)\end{tabular} & $\kappa$ & \begin{tabular}[c]{@{}c@{}}$\sigma_{\rm NRDM}^{\rm SI}/m_{\rm NRDM}$\\ (cm$^2$/GeV)\end{tabular} \\ \hline
XENONnT\,\cite{XENON:2020kmp}    & 4                                                         & 50                                                        & 0.29     & $1.66 \times 10^{-50}$                                                                         \\ \hline
LZ (future)\,\cite{LUX-ZEPLIN:2018poe} & 6                                                         & 30                                                        & 0.16     & $1.98 \times 10^{-50}$                                                                         \\ \hline
Darwin\,\cite{Schumann:2015cpa,DARWIN:2016hyl}      & 5                                                       & 35                                                      & 0.20   & $3.04 \times 10^{-51}$                                                                         \\ \hline
\end{tabular}
\end{center}
\caption{Required details to obtain the future limit on $\sigma_{\chi}^{\rm SI}$ given in Eq.\,\eqref{eq:CRDMproj}.}
\label{tab:numdetails}
\end{table}

The recoil nuclear energy region of operation for each of the considered experiments is given in Table\,\ref{tab:numdetails}. For NRDM, from the projected limits of XENONnT\,\cite{XENON:2020kmp}, LZ\,\cite{LUX-ZEPLIN:2018poe}, and Darwin\,\cite{Schumann:2015cpa, DARWIN:2016hyl}, we estimate the numerical values of $\sigma_{\rm NRDM}^{\rm SI}/m_{\rm NRDM}$ at higher NRDM mass. Note that $\sigma_{\rm NRDM}^{\rm SI}/m_{\rm NRDM}$ remains constant beyond $m_{\rm NRDM} \gtrsim 100$ GeV, we quote the corresponding values in Table\,\ref{tab:numdetails}.

In Fig.\,\ref{fig:limit}, we show the future projections of XENONnT, LZ, and Darwin, derived in this work, by the dashed orange, purple, and blue contours, respectively. The projected JUNO limit is shown by the dot-dashed violet line, adapted from Ref.\,\cite{Cappiello:2019qsw}. The lower limits are obtained using Eq.\,\eqref{eq:CRDMproj} with the Global fit model given in Ref.\,\cite{Gaisser:2013bla} as the input flux (mentioned in Sec.\,\ref{sec:CRDM}).\footnote{We have matched our result with DarkSUSY, by implementing Global fit CR flux model for CRDM in DarkSUSY\,\cite{Bringmann:2022vra}.}  Expectedly, compared to the current LZ result, currently running XENONnT or LZ experiments would be able to probe factor $\sim 3$ smaller cross-section even for sub-GeV DM. Note that recent Super-K result rules out most of the region of the parameter space that can be probed by future xenon-based experiments\,\cite{Super-Kamiokande:2022ncz}. Further, upcoming Darwin experiment will be able to probes a factor $\sim 10$ smaller DM-nucleon scattering cross section.

\section{Conclusions}
\label{sec:con}
The quest for particle DM through DD experiments has achieved unprecedented background suppression which allows us to probe various well motivated DM models. This mainly revolves around the search for weak-scale DM-nuclear scattering. However, the growing interest in the light-dark sector has triggered the search for light DM ($m_{\chi} \lesssim 1$ GeV) particles in DD experiments. This can be done broadly by lowering the threshold of the detector, changing the target, or boosting the ambient DM particles. We focus on the CR boosted DM scenario. Particularly we explore the prospect of detecting cosmic ray boosted sub-GeV DM in light of recent LZ results and future xenon experiments.

With the current LZ result, in the absence of any DM signal, we set the leading constraint on SI DM-nucleon scattering cross-section for sub-GeV DM. Using the parametric form of CR flux as the input, we show that the latest LZ result sets a factor $\sim 2$ stringent constraint compared to the same XENON1T limit for DM mass near 1 MeV. We also find that for XENONnT/LZ(future) and Darwin, there would be factors of $\sim 3$ and $\sim 10$ improvements, respectively. This would allow for complementary probe of SI DM-nucleon scattering cross-section.

\paragraph*{Acknowledgements\,:} TNM thanks IOE-IISc fellowship program for financial assistance. RL acknowledges financial support from the Infosys foundation (Bangalore), institute start-up funds, and Department of Science and Technology (Govt. of India) for the grant SRG/2022/001125.


\bibliographystyle{JHEP}
\bibliography{ref.bib}

\end{document}